**Spin qubit based on the nitrogen-vacancy center analog in a diamond-like compound C$_3$BN**

Duo Wang, Lei Liu, and Houlong L. Zhuang

School for Engineering of Matter, Transport and Energy, Arizona State University, Tempe, AZ 85287, USA

*zhuanghl@asu.edu

**Abstract**

The Nitrogen-vacancy (NV) center in diamond plays important roles in emerging quantum technologies. Currently available methods to fabricate the NV center often involve complex processes such as N implantation. By contrast, in a diamond-like compound $C_3BN$, creating a boron (B) vacancy immediately leads to an NV center analog. We use the strongly constrained and appropriately normed (SCAN) semilocal density functional—this functional leads to nearly the same zero-phonon line (ZPL) energy as the experiment and as obtained from the more time-consuming hybrid density functional calculations—to explore the potential of this NV center analog as a novel spin qubit for applications in quantum information processing. We show that the NV center analog in $C_3BN$ possesses many similar properties to the NV center in diamond including a wide band gap, weak spin-orbit coupling, an energetically stable negatively charged state, a highly localized spin density, a paramagnetic triplet ground state, and strong hyperfine interactions, which are the properties that make the NV center in diamond stand out as a suitable quantum bit (qubit). We also predict that the NV center analog in $C_3BN$ to exhibit two ZPL energies that correspond to longer wavelengths close to the ideal telecommunication band for quantum communications. $C_3BN$ studied here represents only one example of *A₃XY (A:* group IV element; *X/Y:* group III/V elements) compounds. We expect many other compounds of this family to have similar NV center analogs with a wide range of ZPL energies and functional properties, promising to be new hosts of qubits for quantum technology applications. Furthermore, *A₃XY* compounds often contain group IV elements such as silicon and germanium, so they are compatible with the sophisticated semiconductor processing techniques. Our work opens up ample opportunities towards scalable qubit host materials and novel quantum devices.

**Introduction**

Despite in its early infancy, quantum computers hold great potential for solving extremely challenging problems faced by currently available supercomputers, ranging from factorizing a large integer to break public-key cryptography to discovering drugs to treat pandemic diseases [1]. At the heart of the superpowers of quantum computers are the materials that constitute the quantum bits (i.e., qubits) having two energy levels that are analogous to the classical bits represented by 0s and 1s. To realize the full potential of quantum computers in the foreseeable future, advancing the development of building blocks is the key. Among the many proposed candidates for qubit materials such as trapped ions [2] and superconducting circuits [3] that work at near absolute zero, solid-state spin qubits based on defect centers embedded in wide-band-gap semiconductors such as diamond are promising that can lead to novel quantum devices capable of operating at room temperature [4].

The NV center in diamond corresponds to a peculiar defect configuration, where one C atom in a pair of nearest-neighboring (NN) C atoms is removed and the other replaced by a N atom. Although the defect can exhibit different charge states, the NV center commonly refers to the negatively charged state with an extra electron denoted as NV$_{-1}$. This charged state engages the interplay of six electrons, three of which come from the three C dangling bonds surrounding the C vacancy, two from the N atom (the other three electrons of the N atom are shared with its three NN C atoms), and the rest from the donor in the bulk [5]. The total spin resulting from the energy distribution of the six electrons is one and the ground and excited states of the NV center are triplet states with sublevels of $m_s$ = 0, -1, or 1. The $m_s = \pm 1$ states are separable by a small microwave field causing the Zeeman spitting. The superposition of the $m_s = 0$ state and one of the $m_s = \pm 1$ states behaves as the $|0\rangle$ and $|1\rangle$ levels of a qubit, respectively. The initialization and readout of a qubit based on the NV center are realized via optical pumping and spin-dependent fluorescence, respectively [6]. NV center qubits have been shown to exhibit long room-temperature coherence times ($T_2$) on the timescale of microseconds [7], owing to weak spin-orbit coupling and hyperfine interactions [8]. Furthermore, the spins in two NV center qubits have been demonstrated to entangle at room temperature [9], an encouraging step towards scalable quantum register consisting of multiple qubits.

Despite the abovementioned potential of the NV center in diamond has shown in quantum technology applications, there are limitations from the perspectives of manufacturing and zero-phonon line (ZPL) energy. In terms of manufacturing, fabricating these defect centers in a prescribed manner is challenging. Two common methods to create the NV center are (i) chemical vapor deposition (CVD) growing diamond and (ii) high-energy N implantation followed by annealing processes [10]. The locations of the NV center resulted from both methods cannot be accurately manipulated at the atomic scale [6], leading to randomly distributed NV centers and degrading the quality of a quantum register. Furthermore, it is challenging to fabricate devices from diamond [11]. Significant amounts of experimental and theoretical efforts have been spent to identify defect centers in other wide-band-gap semiconductors such as SiC polytypes [12] and GaN [13]. In terms of the ZPL energy, it takes up only 3-4% of the total emission [14]. Because future quantum computers based on the NV center qubits will communicate through optical fiber, enhancing the ZPL wavelengths so that they can match the ideal telecommunication band wavelength is important to minimize the optical loss.

We aim to search for defect centers for hosting spin qubits from a new group of functional semiconductors with a common formula $A_3XY$, where $A$ is a group IV element, $X$ and $Y$ are groups III and V elements, respectively. We expect $A_3XY$ to be free of the limitations posed by diamond and meanwhile to possess the potential as novel qubit materials for three reasons. First, as illustrated in Figure 1(a)-(c), the structure of an $A_3XY$ compound consists of tetrahedral molecular geometry similar to that in diamond, although the former geometry does not have the $C_{3v}$ symmetry as the latter. Each $X$ or $Y$ atom has three nearest neighboring $A$ atoms, so the overall structure can be regarded as embedding $XY$ diatomic units in diamond. $X$ and $Y$ atoms have eight valence electrons, so the average number of valence electrons per atom in $A_3XY$ is the same as in diamond. The special atomic arrangement as well as the isoelectric feature facilitate the formation of an NV center analog. In other words, removing one $X$ atom and negatively charge it with an extra electron is akin to creating a charged $Y$-vacancy center in $A_4$. Second, the mature semiconductor industry is well prepared for new materials that contain group IV elements such as silicon and germanium. $A_3XY$ compounds lie in this category, so the quantum devices made of $A_3XY$ are most likely compatible to the existing semiconductor techniques and scalable to integrate a significant number of qubits. Third, one notable difference of $A_3XY$ from pure bulk $A_4$ is that $A_3XY$ have mixed non-polar and polar bonds, which are expected to affect the ZPL wavelengths via the presence of diverse band gaps and functional properties.

Synthesizing $A_3XY$ using conventional CVD method is nevertheless challenging and subject to the possibility of phase separation. Kouvetakis et al. overcame this challenge via molecular beam epitaxy based on a gas source and successfully synthesized a prototype $A_3XY$ compound $Si_3AlP$. Recent theoretical studies have predicted $Si_3AlP$ to have a direct band gap that is potentially helpful for solar cell applications due to its direct band gap [15]. It was also suggested that the same experimental approaches could be applied to synthesize many other $A_3XY$ compounds [16].

In this work, we focus on one representative example of $A_3XY$ compounds, i.e., $C_3BN$. We explore the potential of $C_3BN$ as a novel semiconductor to host spin qubits. Experimentally,

Langenhorst and Solozhenko synthesized this compound via a shock-compression method at high pressure and temperature [17]. Theoretically, the crystal structure of $C_3BN$ and structural properties such as bulk modulus were first reported in Ref. [18] via density functional theory (DFT) calculations. A more recent DFT study predicts that the hardness of $C_3BN$ is comparable to diamond [19]. As mentioned above, if $C_3BN$ can be successfully synthesized, removing a B atom in $C_3BN$ is equivalent of creating a NV center analog and the whole $C_3BN$ structure has the same number of electrons per atom as diamond. Because the majority atoms in are C and the properties should be similar to those of diamond. Here we apply DFT calculations to compute the properties that are deemed as required ones for a charge defect in a semiconductor to be a qubit candidate. These properties include the electronic structure of bulk $C_3BN$, the defect formation energy, electronic structure, defect energy levels, and hyperfine tensors of the NV-center analog. We compute the same properties of diamond and the NV center for a throughout comparison. We show that the NV center analog in $C_3BN$ satisfies all the criteria for a defect center to become a candidate of qubit materials, endowing $C_3BN$ with a promising potential for hosting spin qubits.

**Methods**

We use the Vienna *Ab initio* Simulation Package (VASP) for all the DFT calculations [20,21]. The potential datasets for the C: $2s^22p^2$, B: $2s^22p^1$, and N: $2s^22p^3$ valence electrons are from the Projector augmented-wave method [22,23]. In all the VASP calculations, the plane waves cutoff kinetic energy is set to 400 eV. For the exchange-correlation interactions, we use the strongly constrained and appropriately normed (SCAN) semilocal density functional, which is one of the most recently developed meta-GGA functionals developed by Perdew and coworkers [24]. This functional not only includes electron density but also local kinetic energy density in the exchange-correlation approximations. SCAN is known to satisfy all the 17 known exact constraints for a semi-local functional. Although SCAN has been applied to many semiconductors to predict their band gaps [25], it has not yet been applied to study the defect levels of the NV center. This work therefore also serves a first example to test the accuracy of this functional in describing the NV center.

To obtain the ground-state structure of $C_3BN$, we adopt the optimized structure of $Si_3AlP$ reported in Ref. [15] as the starting structure for geometry optimizations. The simulation cell contains 12 C, 4 B, and 4 N atoms. Symmetry analysis on the optimized structure shows that the crystal structure of $C_3BN$ belongs to the monoclinic Bravais lattice with the space group of *Cc* (group number: 9). In the geometry optimizations of $C_3BN$ and diamond, all the six lattice parameters and atomic coordinates are fully optimized until the force threshold of 0.01 eV is achieved. The Monkhorst-Pack *k*-point grids [26] for these two systems are $4 \times 4 \times 7$ and $8 \times 8 \times 8$, respectively. The optimized lattice parameters of $C_3BN$ are $a = 5.669$ Å, $b = 5.670$ Å, $c = 3.595$ Å, $\alpha = \beta = 90°$, $\gamma = 89.755°$ and the optimized lattice parameters of diamond are $a = b = c = 3.549$ Å and $\alpha = \beta = \gamma = 90°$. The symmetry of $C_3BN$ is much lower than diamond as can be seen from the tetrahedral molecular geometry illustrated in Figure 1(c). We calculate the energies of the 36-atom unit cell of B and the 2-atom cell of a $N_2$ molecule placed in the center of a vacuum box. These two energies will be used to calculate the defect formation energy (see below). The *k*-point grids of these two calculations are $10 \times 10 \times 4$ and $1 \times 1 \times 1$ (Γ point), respectively. We compute the dielectric constant tensor also using the 20-atom and 8-atom cells for $C_3BN$ and diamond, respectively. The dielectric constant tensor is calculated based on the density functional perturbation theory (DFPT) [27]. Because the SCAN version of DFPT is not yet implemented in VASP, we switch to the Perdew-Burke-Ernzerhof (PBE) functional [28] with the optimized geometry using the SCAN functional to compute the dielectric-constant tensor, where local field effects are accounted for on the Hartree level. We calculate the band structure of C$_3$BN and diamond using their primitive cells that have 10 and 2 atoms, respectively. The corresponding Monkhorst-Pack $k$-point grids are $8 \times 8 \times 6$ and $12 \times 12 \times 12$, respectively. We consider the spin-orbit coupling (SOC) [29] in the band structure calculations.

To simulate the NV-center analog in C$_3$BN, we first create a $2 \times 2 \times 3$ supercell of C$_3$BN (240 atoms) from the optimized 20-atom cell so that the three lattice vectors have nearly the same length, and then we remove one B atom whose crystalline coordinates are (0.351, 0.451, 0.499) and we add an extra electron to the supercell. For the simulations of the NV center in diamond, we use a $3 \times 3 \times 3$ supercell (216 atoms) and remove one C atom positioned at (0.500, 0.333, 0.500) and substitute one of its NN C atoms by a N atom located at (0.583, 0.417, 0.583). For the geometry optimizations of these two supercells, we use a single $k$ point ($\Gamma$ point) and optimize only the atomic coordinates while keeping the lattice constants fixed. For the calculations of density of states, spin density, and hyperfine structure, we use a $2 \times 2 \times 3$ $k$-point grid.

**Results and Discussion**

We start with computing the band structures of C$_3$BN and diamond without defects. The band structure of diamond is used for comparison and also for benchmarking the accuracy of the SCAN functional in describing band gaps. Weber et al. summarized nine conditions that a viable semiconductor to host spin qubits [11]. Two of them can be evaluated from the band structure of the candidate, i.e., it must have a wide band gap and weak SOC. The former criterion is required to accommodate deep defect levels; the latter is preferred to avoid disturbing the electron spin and thereby maintaining the long coherence time. Note that strong SOC is not always an unfavorable characteristic for qubits. As a matter of fact, it becomes a desirable property in some other types of qubit host materials such as semiconductor nanowires [30].

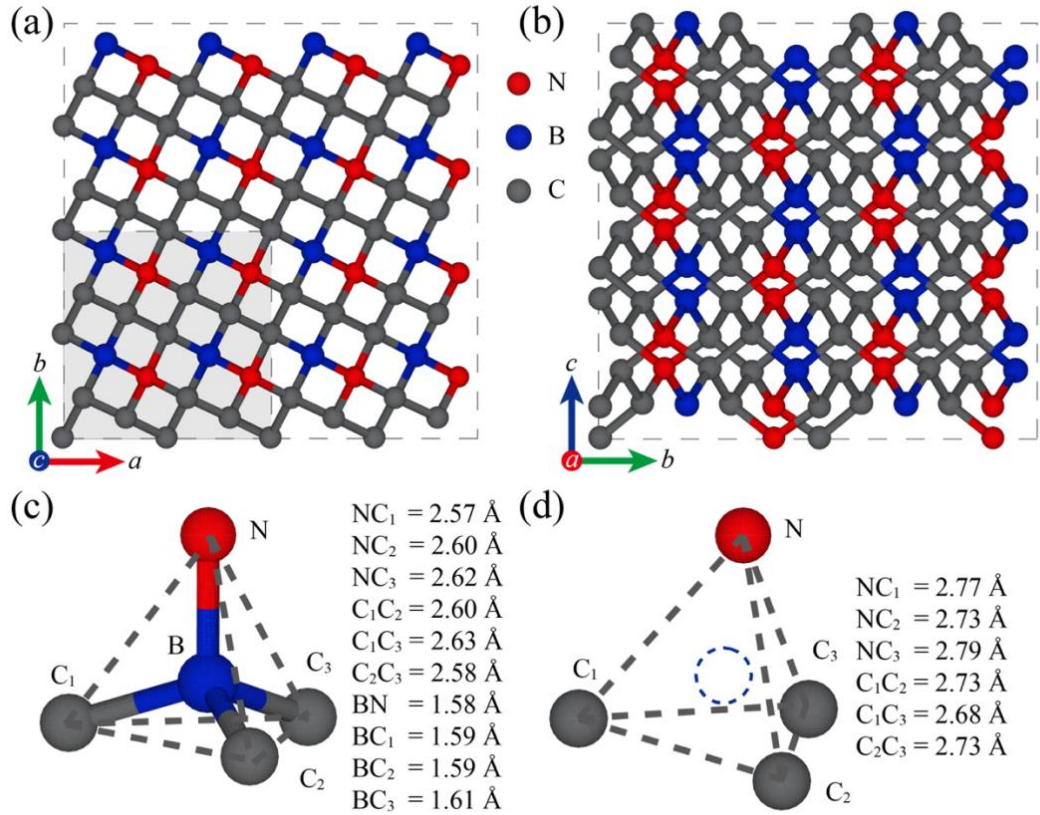

Figure 1. (a) Top and (b) side views of the atomic structures of C₃BN as an example of $A_3XY$ compounds. Here, $A$ = C, $X$ = B, and $Y$ = N. A unit cell of C₃BN is enclosed in the shaded area of (a). Optimized tetrahedral geometries of C₃BN (c) without and (d) with the B vacancy (denoted by the dashed circle). The interatomic distances are shown for both tetrahedral geometries.

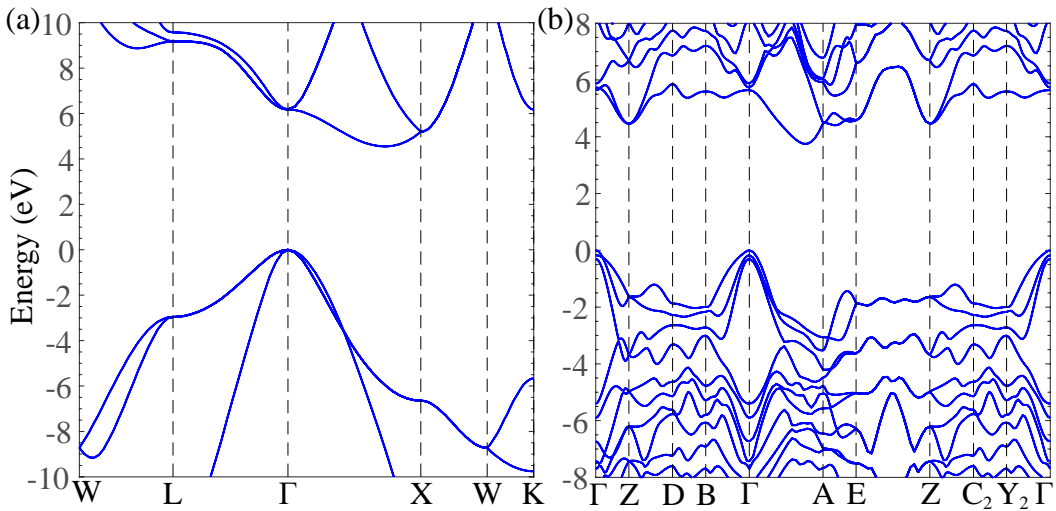

Figure 2. Band structures of (a) diamond and (b) C₃BN calculated with the SCAN functional and spin-orbit coupling is taken into account. The notations of special $k$ points are adopted from Ref. [31]. The valence band maxima are set to zero.

We evaluate the feasibility of $C_3BN$ as a novel semiconductor for hosting spin qubits using the two criteria. Figure 2(a) shows the band structure of diamond with SOC taken into account. We can see that diamond has an indirect band gap of 4.55 eV. Although the SCAN band gap remains underestimating the experimental band gap of 5.50 eV [32], the SCAN functional improves the band gap in comparison to the PBE band gap calculated to be 4.14 eV. The calculated SCAN and PBE band gaps of diamond are both consistent with previous work [25]. Using the SOC splitting at the valence band maximum (VBM) as a metric of SOC strength [5], we find this is negligibly small for diamond. Figure 2(b) shows that $C_3BN$ exhibits an indirect band gap of 3.75 eV, which is smaller than that of diamond, with the VBM at the $\Gamma$ point and the conduction band located between the $\Gamma$ and A points. Furthermore, the top two valence bands at the $\Gamma$ point are nearly degenerate, indicating again a negligible SOC splitting and therefore trivial SOC. Although more advanced theory such as many-body $G_0W_0$ calculations [33] may correct the band gap of $C_3BN$ to even larger numbers, the SCAN band gap is already wide enough to hold deep defect states. Together with the weak SOC, $C_3BN$ seems promising to be another wide-band-gap semiconductor to host spin qubits.

Figure 2(c) shows the tetragonal molecular motif in the optimized structure of $C_3BN$. Most of the interatomic distances are not equal and greater than the NN C-C bond length of 1.536 Å in diamond, revealing the low symmetry of this motif as well as of the bulk. The structure of the motif belongs to the $C_1$ point group. As can be seen in Figure 2(d), creating the NV-center analog leads to enlarged interatomic distances and the point group remains the same.

Although not explicitly listed as one of the nine criteria by Weber et al., the stability of charged defects in a semiconductor is important and needs to be examined. Specifically, as a metric of stability, the defect formation energy provides not only the equilibrium concentration of a charged defect [11] but also the range of Fermi energies (within the limit of the band gap) [34], where a charged state can be more stable than the other possible charged states. We therefore evaluate the stability of charge defects ($q$ = 0, -1, or -2) in C$_3$BN by computing the defect formation energy $E^f\left[C_3BN:NV^q\right]$ following the commonly used equation [35]:

$$E^f\left[C_3BN:NV^q\right] = E_{tot}\left[C_3BN:NV^q\right] - E_{tot}\left[C_3BN:bulk\right] + \mu_B + q\left(\varepsilon_F + \varepsilon_{VBM}^{bulk} + \Delta V\right), \quad (1)$$

where $E_{tot}\left[C_3BN:NV^q\right]$ and $E_{tot}\left[C_3BN:bulk\right]$ are the total energies of the C$_3$BN supercells without and with a charged defect, respectively. $\mu_B$ is the chemical potential of bulk B. $\varepsilon_F$ is the Fermi energy with reference to the VBM $\varepsilon_{VBM}^{bulk}$ of the bulk. $\Delta V$ is the energy correction term due to the energy alignment of the VBM [36]. For comparison, we also compute the formation energy of charged defect in diamond $E^f\left[C:NV^q\right]$ written in a similar equation:

$$E^f\left[C:NV^q\right] = E_{tot}\left[C:NV^q\right] - E_{tot}\left[C:bulk\right] - \mu_N + 2\mu_C + q\left(\varepsilon_F + \varepsilon_{VBM}^{bulk} + \Delta V\right), \quad (2)$$

where $E_{tot}\left[C:NV^q\right]$ and $E_{tot}\left[C:bulk\right]$ are the total energies of the diamond supercells without and with charges, respectively. $\mu_N$ is the chemical potential of N taken as half of the energy of a N$_2$ molecule. A factor of 2 in Eq. 2 is because of the two missing C atoms to create the NV center. $\mu_C$ is the chemical potential of C in bulk diamond, namely, $\mu_C = E_{tot}\left[C:bulk\right]/216$.

The energy correction term $\Delta V$ in Eqs. (1) and (2) also depends on the dielectric-constant tensors of C$_3$BN and diamond. We therefore calculate the dielectric constants in C$_3$BN whose $\varepsilon_{aa}$, $\varepsilon_{bb}$, and $\varepsilon_{cc}$ components are 5.51, 5.75, and 5.52, respectively. The other three components are almost zero. For comparison, the computed dielectric constant of diamond, 5.50, is consistent with the experimental dielectric constant 5.68 [37], showing the accuracy of the SCAN functional. We notice that the dielectric constant components of C$_3$BN in the *a* and *c* directions are nearly the same as those in diamond, manifesting the high similarity between C$_3$BN and diamond in these two directions. Figure 1(a) shows that along the positive *a* direction, B and N atoms appear to locate in zigzag chains, each of which can be denoted as (BN)(BN)(BN). Because each diatomic (B and N) unit is close in distance, the pattern can also be regarded as (NB)(NB)(NB) along the negative *a* direction. These two different plausible denotations imply that the polarization due to the BN polar bonds along the positive and negative *a* directions will cancel out, leading to almost the same dielectric constant with diamond in the *a* direction. Similarly, Figure 1(b) shows that along the positive or negative *c* direction, B and N atoms always follow the same (BN)(NB)(BN)(NB)(BN)(NB) pattern. As a result, the polarization in the BN bonds plays little role in affecting the dielectric constant in the *c* direction. By contrast, along the positive and negative *b* directions, the distance between the diatomic unit is so large that the pattern can only be regard as (BN)(BN)(BN) or (NB)(NB)(NB). The dielectric constants of C$_3$BN in the *b* direction is therefore slightly larger than diamond due to the polar nature of B-N bonds.

In computing the defect formation energy, we consider the neutral state, $q = 0$, and two charged states, $q = -1$ and -2. When $q = -1$, the state refers to the NV-center analog in C$_3$BN or the NV center in diamond. We consider the $q = -2$ state to identify the upper bound of the Fermi energy where the NV-center analog or NV center remains more stable. Figure 3 shows the defect formation energy of C$_3$BN and diamond as a function of the Fermi energy with reference to the VBM. Our computed defect formation energy for the neutral defect in diamond is 5.92 eV, which is comparable to 6.21 eV using a 511-atom supercell [38] and to ~6.10 eV using a 63-atom supercell [11] with the HSE06 functional [39]. Figure 3(a) also shows that the lower bound of the Fermi energy is 1.78 eV, above which the NV center in diamond is more stable than the other two defect states. This lower bound is smaller than ~2.70 and 2.78 eV in Refs. [38] and [11], respectively, possibly due to the different functional used in the current calculations. Our calculations cannot provide an upper bound for the Fermi energy in diamond because of the underestimated band gap by the SCAN functional. However, we consistently show that the NV center is stable in a wide range of Fermi energies. Figure 3(b) shows the formation energies of the three charge states of $C_3BN$. For the neural defect, the formation energy is 6.82 eV in $C_3BN$, higher than that (5.92 eV) in diamond, implying that it is relatively more energy-consuming to create a B vacancy in $C_3BN$ than to generate two C vacancies, one of which is further replaced by a N atom. More importantly, the NV-center analog is the most stable among the three defect states, if the Fermi energies are within the energy range of 1.52 to 3.26 eV. The lower bound is nearly in the middle of the band gap and comparable to that of diamond, indicating that the defect state is deep enough to trap an extra electron. Although the HSE06 functional may give a more accurate band gap of $C_3BN$ and then more accurate lower and upper bounds of Fermi energies, we expect our conclusion that the NV-center analog as a deep charged state to remain the same, according to our results using the SCAN functional.

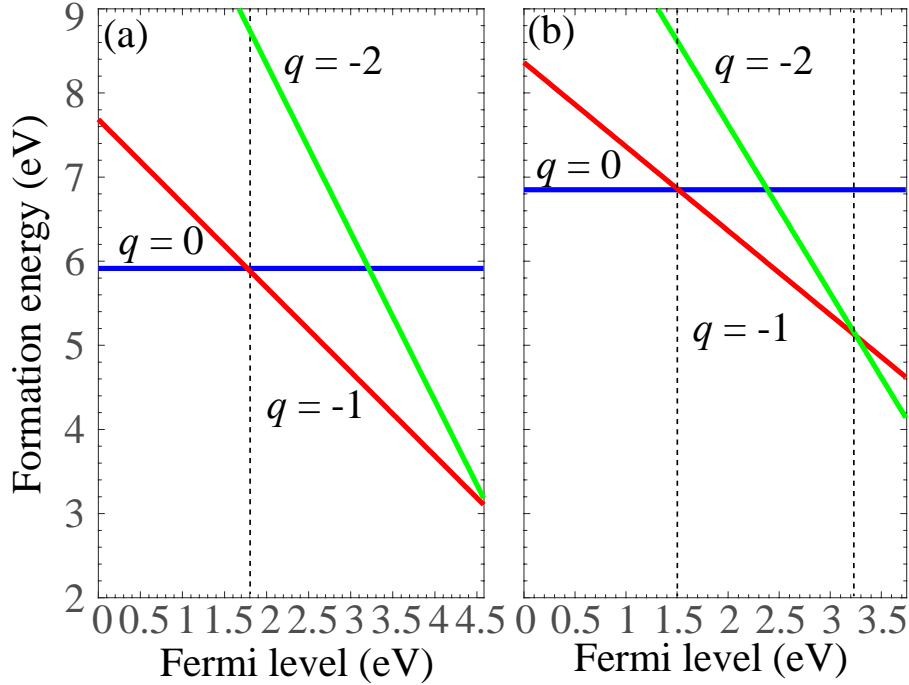

Figure 3. Defect formation energy of (a) diamond and (b) C$_3$BN. Only one dashed line is shown to represent the lower bound in diamond, where the $q = -1$ state is the most stable due to the underestimated band gap with the SCAN functional. The region where the $q = -1$ state in C$_3$BN is the most energetically stable is enclosed by two dashed lines.

Having shown the stability of the NV center analog in C$_3$BN, we now focus on studying the electronic structure of this analog. Figure 4 shows the spin density of states (SDOS) and energy levels of the defect states of the NV center and its analog. For the NV center, its electronic structure has been well studied using different levels of theory. Reference [40] provides an excellent overview this topic. Interestingly, we notice that the SCAN functional has not yet been applied to study the NV center. We show here that using the SCAN functional can reproduce all the key features in the electronic structure of the NV center. First of all, our calculated SDOS agrees well with the SDOS obtained from the PBE functional [41]. The spin-up and spin-down DOS curves are nearly symmetrical except in the band gap where we observe four SDOS peaks. The two spin-up DOS peaks are located at the energies of around -0.273 and -1.443 eV, respectively. By computing the integrated DOS, we find that the higher-energy DOS is occupied by two degenerate electrons often labeled as the $e_x$ and $e_y$ states [5]. Consistent with the smaller area enclosed below the peak, the lower-energy DOS is occupied by only one electron labeled as the $a_1(2)$ state. The two spin-down DOS peaks are located at the energies of about -0.562 and 1.685 eV, respectively. Only the lower-energy DOS is occupied by one electron with the same label, $a_1(2)$. The higher-energy spin-down DOS peak has the same labels as their spin-up counterparts, $e_x$ and $e_y$. This peak is capable of hosting two degenerate electrons. The remaining two electrons of the NV center are embedded in the valence bands. The net effect of the energy distribution of the six electrons is the paramagnetic state with a total spin moment of one Bohr magneton. As described in the Introduction section, this paramagnetic spin is the key factor that makes the NV center a qubit that can be manipulated by magnetic and optical methods.

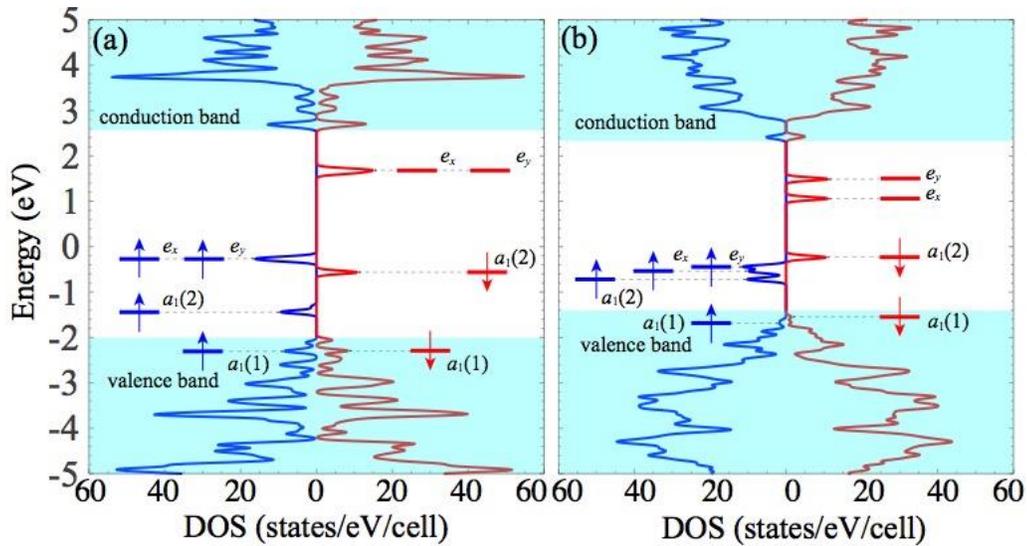

Figure 4. Spin density of states of (a) diamond with the NV center and of (b) B$_3$CN with the NV center analog. Occupied and empty energy levels are represented by line segments with and without overlapped upward (spin-up) and downward (spin-down) arrows, respectively. The upper and lower shaded areas represent the conduction and valence bands, respectively.

Similar to diamond, the SDOS curves shown in Figure 4(b) are also almost symmetric except in the band gap, where we can see six instead of four peaks in diamond. The increased number of peaks is a result of the lowered symmetry in comparison to diamond. Furthermore, two of the six electrons are immersed in the valence band, so they are as well of no significant relevance. The other four electrons in the gap are distributed differently from they are in the NV center. In particular, we notice that the $e_x$ and $e_y$ electrons in the spin-up channel are no longer degenerate and the $a_1(2)$ electron (We use the same notations only for the convenience of comparison.) in the same channel now has a closer energy to the other two spin-up electrons. The three electrons interact to some extent forming a continuous band of width about 0.53 eV. In the spin-down channel, one electron occupies the $a_1(2)$ energy level and the degeneracy of the empty $e_x$ and $e_y$ states is broken separating into two levels with an energy spacing of around 0.43 eV. The total spin moment of the NV center analog is the same as the NV center. This paramagnetic triplet state satisfies another criterion [11] to endow the NV center analog with the potential of hosting a new spin qubit.

Figure 5 compares the spin densities of the NV center in diamond and the NV center analog in $C_3BN$. As can be seen, the spin densities in both the NV center and its analog are highly localized in the C atoms near the vacancy, manifesting another similarity between the $C_3BN$ and diamond. Localized states are often described better by the HSE06 functional than by the PBE functional [42,43]. Our results show that the SCAN functional can also capture this localization very well. The localized spin densities in NV center and its analog form bound states [11], ensuring them to behave as an "artificial atom" that is well protected from the environment and subject to convenient initialization, measurement, and readout.

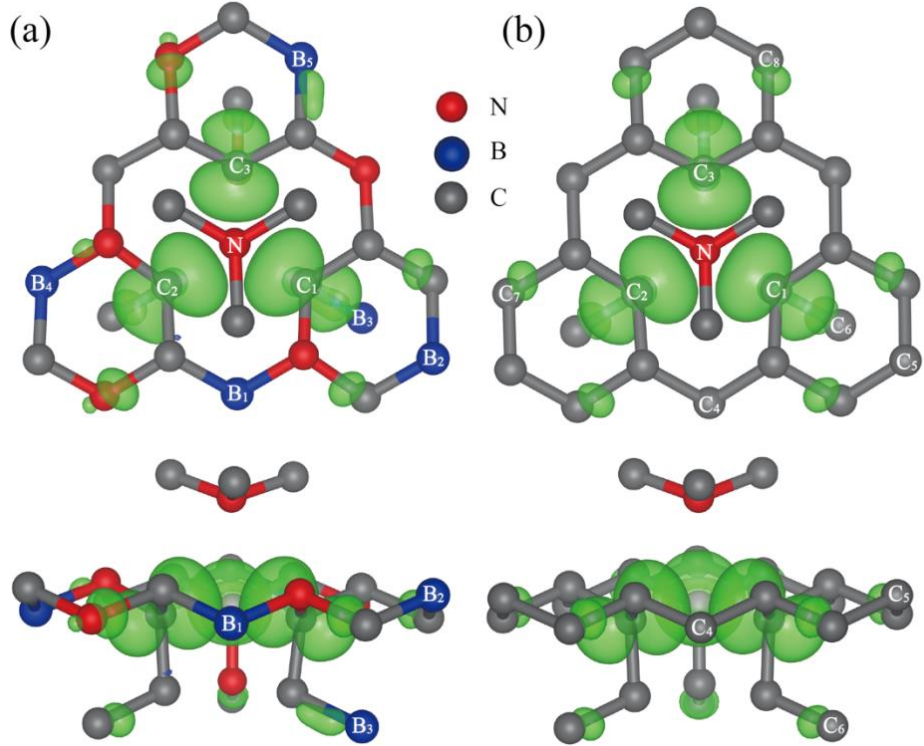

Figure 5. Spin density of (a) the NV center in diamond and (b) the NV center analog in $C_3BN$. The isosurface value is set to 0.005 $e/a_0^3$, where $a_0$ is the Bohr radius.

Because photoluminescence is the main optical approach to manipulate a paramagnetic spin qubit, three of the nine criteria summarized by Weber et al. are related to light-qubit interactions [11]. Over several decades of complimentary experimental and theoretical studies [44,45], a somewhat complete description of light-NV center interactions with fine details such as the non-radiative intersystem crossing is now available for the NV center in diamond. By contrast, the NV center analog in $C_3BN$ has not yet been fabricated, we therefore focus on understanding the spin-conserving radiations based on the energy diagrams shown in Figure 4. In particular, we predict the potential energy as function of configuration coordinate of the NV center analog in $C_3BN$, following the Frank-Condon approximation [46,47]. From the potential energy curves, we extract the critical information such as the ZPL energy that could be confirmed in future photoluminescence measurements. Owing to the broken degeneracy of the two spin-down excited states $e_x$ and $e_y$ in the NV center analog in $C_3BN$, there are two possible cases of spin-conserving optical transitions: case 1, transition from the spin-down $a_1(2)$ state to the spin-down $e_x$ state; case 2, transition from the spin-down $a_1(2)$ state to the spin-down $e_y$ state. We thus expect to observe two ZPL peaks for the NV center analog in $C_3BN$. We predict the ZPL energies and other states using the constrained DFT [48] as implemented in VASP. Figure 6 shows the sketches of potential energy curves for the NV center in diamond and for the two cases of the NV center analog in $C_3BN$. The curves for each system involved four states A, B, C, and D. State A is the ground state of the NV center or its analog; state B is the energy of placing the spin-down $a_1(2)$ electron at the spin-down $e_x$ or $e_y$ state without atomic relaxation; state C is the energy reduction from optimizing the structure of state B and this energy reduction corresponds to the Stokes shift (S). Placing the electron back to the $a_1(2)$ state followed by geometry optimizations leads to state D, whose energy is less than that of state C by an amount of the anti-Stock-shift (AS).

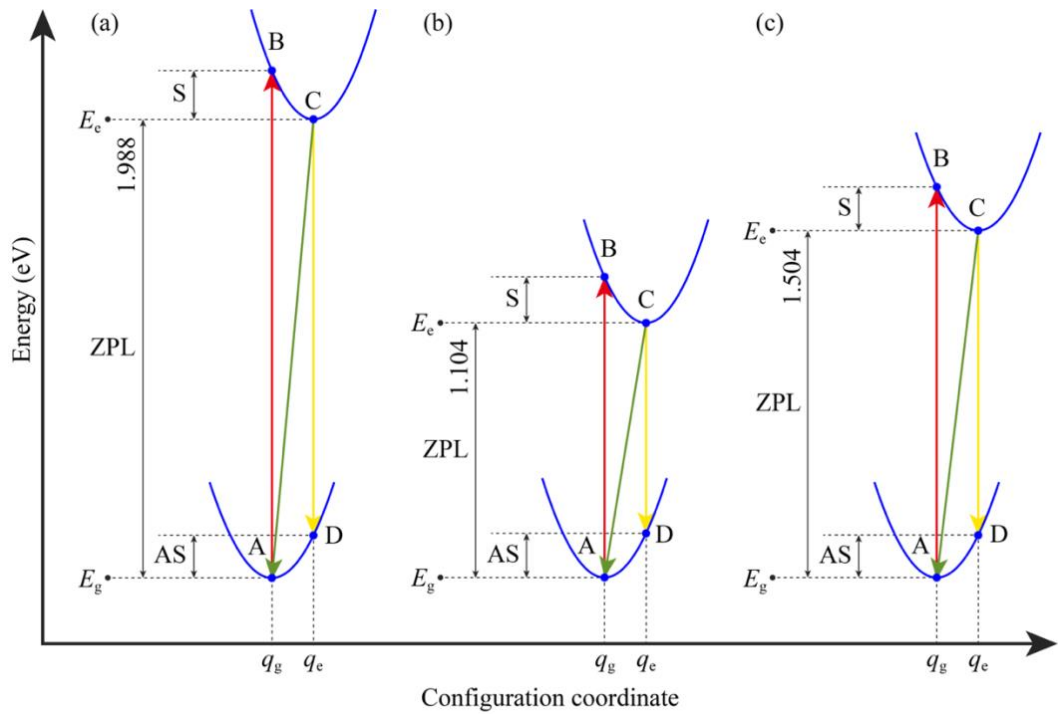

Figure 6. (a) Sketched potential energy curves of the NV center in diamond and of the NV center analog with (b) case 1 and (c) case 2 transitions.

Table 1 lists the ZPL energies, A→B and C→D transition energies, and the S and AS energies. We notice that these data for the NV center in diamond calculated using the SCAN functional is in excellent agreement with the results obtained from using the more time-consuming HSE06 functional [42] and also with the experimental data [49]. These promising results indicate that it is worthwhile using the SCAN functional to study other emerging candidates of point defects such as divacancy in 4H-SiC [50] for hosting spin qubits. Figure 6 and Table 1 reveal an attractive property of the NV center analog in $C_3BN$. That is, the ZPL energies in the two cases are much smaller than that of the NV center in diamond. These lower excitation energies, corresponding to the significantly longer photon wavelengths of 1123 and 824 nm, respectively, which are closer to the ideal telecommunication band between 1310 (the "O-band") and 1550 nm (the "C-band") [51]. We therefore expect the NV analog in $C_3BN$ to be advantageous over the NV center in experiencing less optical loss in optical fiber that connects qubits to form a quantum network.

Table 1. Zero-phonon line (ZPL) energy, A→B and C→D transition energies, and the Stokes (S) and anti-Stokes-shift (AS) of diamond and $C_3BN$ computed with the SCAN functional. The notations are shown in Figure 6. Many data for the NV center in diamond are available. We show here two representative references for comparison: one from using the HSE06 hybrid density functional, the other from experimental data.

|                      | ZPL   | A→B   | S     | C→D   | AS    |
|----------------------|-------|-------|-------|-------|-------|
| Diamond[1]           | 1.988 | 2.199 | 0.211 | 1.804 | 0.184 |
| Diamond[2]           | 1.955 | 2.213 | 0.258 | 1.738 | 0.217 |
| Diamond[3]           | 1.945 | 2.180 | 0.235 | 1.760 | 0.185 |
| $C_3BN$[4]           | 1.104 | 1.303 | 0.199 | 0.912 | 0.192 |
| $C_3BN$[5]           | 1.504 | 1.693 | 0.189 | 1.320 | 0.184 |

[1]This work using the SCAN functional
[2]Ref. [42] using the HSE06 functional
[3]Ref. [49]; experimental data
[4]This work, case 1 transition, using the SCAN functional
[5]This work, case 2 transition, using the SCAN functional

Finally, we evaluate the hyperfine structure of the NV center and its analog. The hyperfine interactions between the nuclear spin and the qubit spin are the main source that lead to the decoherence of spin qubits [52]. The sources of hyperfine interactions in diamond are isotopes $^{13}$C and $^{14}$N with nuclear spins of 1/2 and 1, respectively. An extra source in C$_3$BN is isotope $^{11}$B with a nuclear spin of 3/2. Due to the high chance of presence of $^{13}$C in the NV center [53], the hyperfine interactions have recently been employed for quantum error correction [54] and entanglement distillation [55]. In this context, stronger hyperfine interactions become a preferable property. In experiment, two of the three C atoms are $^{13}$C and the third one is $^{12}$C with no nuclear spin [56,57]. The two nuclear qubits along with the one spin qubit form a three-qubit register. We here consider an extreme scenario, where all the three C atoms near the vacancy are isotope $^{13}$C. We calculate the hyperfine tensors that include the Fermi contact and dipole-dipole coupling terms [58]. The gyromagnetic ratios of these three isotopes ($^{13}$C, $^{14}$N, and $^{11}$B) are 10.7084 [59], 3.077 [59], and 13.7 [60], respectively, adopted as inputs for the calculations. Table 2 reports the computed hyperfine tensors of the nine atoms near the vacancy site in diamond and C$_3$BN. We can see that the three $^{13}$C atoms exhibit the strongest hyperfine interactions in both systems whereas the $^{14}$N atoms shows negligible hyperfine interactions. The hyperfine tensors of the C atoms in C$_3$BN are comparable and some of them are even higher than their counterparts in diamond, suggesting that a three-qubit register in C$_3$BN can also be employed for quantum error correction and entanglement distillation. Table 2 also shows that the hyperfine tensors for the other C atoms in diamond and the corresponding B atoms in C$_3$BN are smaller by an order of magnitude, as these atoms are relatively further apart from the spin densities. We note that the hyperfine interaction because of $^{11}$B cannot be removed via isotope engineering (unlike C in diamond which can be isotopically purified to have zero-spin $^{12}$C), so we should expect small effects of $^{11}$B atoms on the decoherence of the spin qubit in C$_3$BN.

Table 2. Principal values (in MHz) of the total hyperfine tensors of three C, one N, and five C/B atoms near the vacancy in diamond and C$_3$BN. The notations for these nine atoms are shown in Figure 5. The total hyperfine tensors are calculated using the SCAN functional.

|  | Diamond | | | C$_3$BN | | |
|---|---|---|---|---|---|---|
|  | $A_{xx}$ | $A_{yy}$ | $A_{zz}$ | $A_{xx}$ | $A_{yy}$ | $A_{zz}$ |
| C$_1$ | 147.959 | 147.696 | 231.746 | 172.525 | 171.442 | 263.850 |
| C$_2$ | 147.959 | 147.696 | 231.747 | 150.745 | 150.240 | 234.171 |
| C$_3$ | 147.959 | 147.696 | 231.747 | 135.154 | 134.460 | 205.614 |
| N | -2.445 | -1.939 | -2.445 | -1.125 | -0.475 | -1.204 |
| C$_4$/B$_1$ | 4.041 | 2.457 | 4.128 | 2.398 | 1.258 | 3.489 |
| C$_5$/B$_2$ | -1.846 | -1.003 | -1.869 | -0.888 | 0.101 | -1.039 |
| C$_6$/B$_3$ | 15.589 | 15.447 | 21.106 | 9.016 | 8.804 | 13.564 |
| C$_7$/B$_4$ | 17.003 | 16.908 | 22.918 | 5.267 | 4.997 | 7.855 |
| C$_8$/B$_5$ | 17.003 | 16.908 | 22.918 | 11.51 | 11.203 | 14.88 |

**Conclusions**

To summarize, we have shown that a negative charged ($q = -1$) B vacancy defect in a diamond-like compound C$_3$BN forms an NV center analog that shares a number of common properties to the NV center in diamond. Specifically, the band structure computed from the SCAN functional exhibits a wide band gap of 3.75 eV and negligible SOC. The charged defect state is energetically stable in a wide range of Fermi energies. The ground state of this charged vacancy state has a total spin of 1 that can be manipulated via optical approaches to form a two-level system. These properties make $C_3BN$ a promising semiconductor to host spin qubits made of the NV center analog for quantum information processing. We also showed that the experimental and HSE06 ZPL energies of the NV center in diamond are well reproduced using the SCAN functional. We therefore suggest this functional used for discovering other candidates of qubit host materials. Furthermore, we computed the ZPL energies of the NV center and its analog and found that the ZPL wavelengths of the NV center are longer and closer to the ideal telecommunication band wavelength, indicating less optical loss if the analog is used in a quantum network with multiple qubits. The computed hyperfine structures of the analog and the NV center are comparable and sufficiently strong to form a quantum register beneficial for quantum error correction. Exemplified by $C_3BN$, many other *A*$_3$*XY* compounds and their potential of hosting NV center analogs will be explored is our future work.

**Acknowledgements**

We thank John Kouvetakis at Arizona State University (ASU) for bringing *A*$_3$*XY* compounds to our attention and for help discussion. We also thank the start-up funds from ASU. This research used computational resources of the Texas Advanced Computing Center under Contracts No.TG-DMR170070 and the Agave cluster at ASU.